\documentclass[11pt,a4paper]{article}
\usepackage{amsfonts}
\usepackage{amsmath,amssymb}
\usepackage{epsfig,graphicx}

\input{tcilatex}
\begin{document}

\begin{center}
{\LARGE Deriving Gauge Symmetry and Spontaneous Lorentz Violation}

\bigskip

{\bf J.L.~Chkareuli}$^{1}$, \ {\bf C.D.~Froggatt}$^{2}${\bf \ and \ H.B.
Nielsen}$^{3}$

\bigskip

$^{1}${\it E. Andronikashvili} {\it Institute of Physics and }

{\it I. Chavchavadze Tbilisi State University, 0177 Tbilisi, Georgia\
\vspace{0pt}\\[0pt]
}

$^{2}${\it Department of Physics and Astronomy, Glasgow University,}\\[0pt]
{\it Glasgow G12 8QQ, Scotland}\vspace{0pt}\\[0pt]

$^{3}${\it Niels Bohr Institute, Blegdamsvej 17-21, DK 2100 Copenhagen, }

{\it Denmark}

\bigskip \bigskip

{\bf Abstract}

\bigskip
\end{center}

We consider a class of field theories with a four-vector field $A_{\mu }(x)$
in addition to other fields supplied with a global charge symmetry -
theories which have partial gauge symmetry in the sense of only imposing it
on those terms in the Lagrangian density which have derivatives as factors
in them. We suppose that spontaneous Lorentz invariance breaking occurs in
such a theory due to the four-vector field taking a non-zero vacuum
expectation value. Under some very mild assumptions, we show that this
Lorentz violation is not observable and the whole theory is practically
gauge invariant. A very important presupposition for this theorem is that an
initial condition is imposed on the no-derivative expressions corresponding
to the early Universe being essentially in a vacuum state. This condition
then remains true forever and can be interpreted as a gauge constraint. We
formulate the conditions under which the spontaneous Lorentz violation
becomes observable. Spontaneously broken Lorentz invariance could be seen by
some primordially existing or created \textquotedblleft
fossil\textquotedblright\ charges with the property of moving through the
Universe with a fixed velocity.

\thispagestyle{empty}\newpage

\section{Introduction}

It is by now a rather old idea to seek to obtain the photon as the
Nambu-Goldstone boson for spontaneous Lorentz invariance violation (SLIV)
\cite{bjorken}. If, for instance, a four-vector field $A_{\mu }$ takes on in
vacuum a non-zero temporally and spatially constant value we have an example
of the SLIV \cite{alan,jackiw,cfn}. A priori one would then expect in
theories with such spontaneous breakdown to see in practice violation of
physical Lorentz invariance. It is the purpose of the present article to
point out that under some rather mild assumptions, however, one {\em one
does not find this breaking of Lorentz invariance to be recognizable
experimentally}\footnote{%
Note that previous direct calculations of Lorentz violating processes in the
tree \cite{nambu} and one-loop approximation\cite{ac} for QED with the
non-linear vector field constraint $A^{2}$ $=$ $V^{2}$ (where $V$ is the
SLIV scale, see below), which appear to be completely cancelled, provide
explicit examples of this theorem. On the other hand, many counter examples
with physical Lorentz violation also exist when underlying assumptions for
this theorem are not satisfied\cite{alan,jackiw}.}.

A very simple way to achieve the SLIV is to introduce into a usual QED
Lagrangian density a potential term $P(A^{2})$ for the four-vector field $%
A_{\mu }(x)$ \ (so that $A^{2}(x)=A^{\mu }(x)A_{\mu }(x)$) which then after
having been transferred to the Hamiltonian density in the usual way could
have its minimum for a non-zero value of $A_{\mu }$\cite{alan,cfn}. In this
way a vacuum apparently without Lorentz symmetry can rather easily come
about
\begin{equation}
A_{\mu }(x)=a_{\mu }(x)+n_{\mu }V
\end{equation}%
where $n_{\mu }$ is an properly oriented unit Lorentz vector with $%
n^{2}=n_{\mu }n^{\mu }=\pm 1$ for time-like and space-like Lorentz
violation, respectively, while $V$ is a proposed SLIV scale.

At first one could easily come to the belief that in such models with SLIV
one will find scatterings that indeed reflect the special direction in which
the symmetry is broken and thus that one obtains a theory that for all
practical purposes breaks Lorentz invariance. However, we deliver a theorem
telling that really in a very large class of cases such models {\em do not
show Lorentz symmetry breaking} but rather behave as an ordinary Lorentz \
and gauge invariant theory manifesting the SLIV only in some noncovariant
gauge choice.

In section 2 we shall put up the type of model which we consider: a rather
general type of field theory with a four-vector field $A_{\mu }(x)$ which
although it comes to play the role of the four vector potential in
electrodynamics is a priori {\em not }assumed to obey full gauge invariance
- only terms in the Lagrangian with derivatives are assumed to be gauge
invariant. And we put forward and prove rather closely related versions of
our theorem.

In sections 3 and 4 we argue how it is possible at all to avoid our theorem
first by means of initial conditions with what is essentially immovable
charges, and then by allowing extra terms in the Lagrangian which give no
way to this theorem to work.

In section 5 we make a discussion of that the presuppositions of our theorem
- especially concerning the need for gauge symmetry of the derivative
containing terms - are quite reasonable to require in order to avoid
bottomlessness of the Hamiltonian and/or to avoid domain walls in cosmology.

Finally, in section 6 we present a resume and conclude.

\section{The theorem}

Now we should have in mind that it is a priori the philosophy of the present
article to work with theories that are {\em not} a priori gauge invariant.
This means that we do not impose gauge invariance on the whole Lagrange
density ${\cal L}(x)$. Instead, we shall impose gauge symmetry only on the
vector field kinetic term or, in general, on the part of our Lagrangian
density having derivatives (see for more disussion section 5). Most
importantly this means that the kinetic term for the four-vector field $%
A_{\mu }(x)$ to be the usual one, $-\frac{1}{4}F_{\mu \nu }^{2}$, so as to
see that, in spite of that one would have expected to find effects of
Lorentz breaking (stemming, say, from some vector field constraint $A^{2}$ $=
$ $V^{2}$ where $V$ is the SLIV scale\cite{nambu,ac}), it turns out to
deliver only simple free Maxwell equations. We shall generalize our type of
models to include matter fields such as Dirac fields or Weyl fields and a
charged scalar that could potentially be used as a Higgs field $\phi $ and
also we include into consideration as minimal QED couplings so the
non-minimal ones with dimensionful coupling constants. With such matter
fields our assumption of the terms with derivatives to be gauge invariant
means that most of the terms in the matter field Lagrangian density should
actually have the usual gauge invariant form as it takes place for the
conventional minimal electrodynamics\footnote{%
Note that minimal (dimensionless coupling) vector-fermion field Lagrangian
satisfying the additional global symmetry requirements including charge
conjugation, parity reflection and fermion number conservation automtically
appears as the conventional QED provided the partial gauge symmetry, namely
gauge symmetry of the vector field kinetic term in the Lagrangian is
presupposed.}. However, for the general vector field potential $P(A^{2})$
and the non-minimal couplings in the Lagrangian many new gauge breaking
terms, including the terms with derivatives, could in principle appear in it
through the radiative corrections. We propose that all these terms are very
small being substanially suppressed by the high SLIV scale $V$ which is
usually thought to be close to the Planck mass $M_{P}$. The simplest choice
for possible "large" terms could be then that they were only allowed to
depend on the non-derivative Lorentz invariant field combinations from each
type of field involved $A^{2}$, $|\phi |^{2}$ and $\overline{\psi }\psi $ ($%
\overline{\psi }\gamma _{5}\psi $) separately but not on their mutual
contractions, like as the contraction of the field $A_{\mu }$ with the
fermion current $\overline{\psi }\gamma ^{\mu }\psi $ apart from its usual
occurrence in the minimal QED\footnote{%
Possible operators of dimension 4 or less include a potential term $P(A^{2})$
of fourth order in $A_{\mu }$ and a seagull term $|\phi |^{2}A^{2}$ with an
arbitrary coefficient.}. The complicated high-dimension operators such the
four-fermion current-current term etc. are also ignored. This simplest
choice for gauge breaking terms is then happened to be a base for the
following theorem to work.

\underline{Theorem}: Consider a Lorentz invariant Abelian field theory with
a priori a {\it global }charge conservation symmetry only, while gauge
symmetry is not imposed in as far we allow for terms containing the squared
four vector field $A^{2}$ in combination with globally charge symmetric but
derivative free combinations of the matter fields. This means that the
theory has in addition to only fully gauge invariant terms - as usual
electrodynamics - a (seemingly) gauge breaking term

\begin{equation}
{\cal L}_{br}(A^{2},|\phi |^{2},\overline{\psi }\psi ,\overline{\psi }\gamma
_{5}\psi )  \label{gf}
\end{equation}%
depending only on the globally phase transformation invariant combinations
without any derivatives and only on $A_{\mu }$ via $A^{2}$. Provided now
that the Universe should have started with initial conditions so as to
ensure in early times the vanishing of the partial derivative of the gauge
non-invariant part ${\cal L}_{br}$

\begin{equation}
\frac{\partial {\cal L}_{br}}{\partial A^{2}}=0\quad  \label{gc}
\end{equation}%
and that $A_{\mu }(x)$ equals a non-zero constant in these asymptotic early
times (i.e. a Lorentz symmetry breaking vacuum), then the theory is indeed
interpretable as a Lorentz and gauge invariant theory with a non-covariant
gauge choice properly fixed in the theory.

The basic idea in the proof of this theorem is that once the Universe gets
started in a state in which the \textquotedblright gauge
condition\textquotedblright\ (\ref{gc}) is satisfied, this condition will go
on being fulfilled. In fact, we can easily see that by varying $A_{\mu }$ in
the whole Lagrangian density, which now is of the form

\begin{equation}
{\cal L}={\cal L}_{inv}+{\cal L}_{br}
\end{equation}
where the terms ${\cal L}_{inv}$ are gauge invariant terms the equation of
motion obtained becomes

\begin{equation}
\partial _{\mu }F^{\mu \nu }=2A^{\nu }\frac{\partial {\cal L}_{br}}{\partial
A^{2}}+j_{matter}^{\nu }.  \label{gme}
\end{equation}%
Here $j_{matter}^{\nu }$ is the current coming from the matter fields $\psi $
and $\phi $ other than the four-vector one $A_{\mu }$. With the assumption
of requiring global charge conservation - or phase transformation symmetry -
for the matter fields we get that the current $j_{matter}^{\nu }$ becomes
conserved $\partial _{\nu }j_{matter}^{\nu }=0$, and since the lefthand side
of the equation (\ref{gme}) is divergence free due to the antisymmetry of $%
F_{\mu \nu }$, we thus arrive at the conclusion that the vector field current

\begin{equation}
j_{A}^{\nu }=A^{\nu }\frac{\partial {\cal L}_{br}}{\partial A^{2}}
\label{ja}
\end{equation}
from the non-gauge invariant term becomes itself separately conserved.

Let us first consider the case of the vacuum background $A^{\nu }$ in the
early times were time-like. Then if we start from the mentioned ''gauge
condition'' (\ref{gc}) in early times the conservation of the vector field
current $j_{A}^{\nu }$ (\ref{ja}) comes to say that the partial derivative
of the gauge non-invariant part of the Lagrangian $\frac{\partial {\cal L}%
_{br}}{\partial A^{2}}$ does not vary in the direction of the four vector
field $A_{\nu }$. In fact, the requirement of the separate conservation of $%
j_{A}^{\mu }$ implies that

\begin{equation}
(\partial _{\mu }A^{\mu })\frac{\partial {\cal L}_{br}}{\partial A^{2}}%
+A^{\mu }\partial _{\mu }\frac{\partial {\cal L}_{br}}{\partial A^{2}}=0
\end{equation}%
and thus if $\frac{\partial {\cal L}_{br}}{\partial A^{2}}=0$ at the start,
the derivative of $\frac{\partial {\cal L}_{br}}{\partial A^{2}}$ in the
direction of the the field $A^{\mu }$, namely the second term in this
equation,\ stays equal to zero. So, this partial derivative $\frac{\partial
{\cal L}_{br}}{\partial A^{2}}$ must take the same value all along a curve
laid out to follow the $A^{\nu }$ field direction by having these fields as
tangents. This means that once it is zero at the beginning it must remain
zero along these curves. However, if we start in a vacuum state having
Lorentz invariance spontaneously broken through the vector field $A^{\nu }$
vacuum expectation value ($<A^{\nu }>$ $=n^{\nu }V$ with $n^{2}=1$ for the
time-like SLIV), such a beginning with the \textquotedblleft gauge
condition\textquotedblright\ (\ref{gc}) fulfilled is basically enforced. One
namely then has a non-zero but constant $A^{\nu }$ leading to the $F_{\mu
\nu }=0$. Since the matter current $j_{matter}^{\nu }$ is always zero in
vacuum, it follows that also the current $\ \ j_{A}^{\nu }=0.$Then, with
with $A^{\nu }\neq 0$ we conclude that the factor $\frac{\partial {\cal L}%
_{br}}{\partial A^{2}}$ in the current $j_{A}^{\nu }$ (\ref{ja}) should also
be zero. But once we have $\frac{\partial {\cal L}_{br}}{\partial A^{2}}$
zero at the start, it follows from the equations of motion that this
\textquotedblleft gauge condition\textquotedblright\ is satisfied forever.

So we have seen that a very mild initial condition can enforce the special
gauge condition (\ref{gc}) and the vanishing of the current $j_{A}^{\nu }$
at all times. Thus the potential possibility for seeing e.g. Coulomb fields
in the $F_{\mu \nu }$ around the $j_{A}^{\nu }$ charges, which could reveal
the non-gauge invariant properties of the $j_{A}^{\nu }$ current, is in fact
prevented. That is to say that an observer, who only has access directly to
the usual $F_{\mu \nu }$ fields but not to the $A^{\nu }$ field, could not
hope to gain access to the gauge dependent features of $A^{\nu }$ indirectly
via the $j_{A}^{\nu }$ anymore, because this current would remain zero.

It should be remarked that for the SLIV to be non-observable we had to
assume the special gauge condition (\ref{gc}) as an initial condition, due
to the Universe being at first in a vacuum state. Otherwise, by allowing
initial deviations from the vacuum state, we would get some true observable
breaking of Lorentz invariance. However, such alternative initial conditions
means that there are a kind of \textquotedblright fossils\textquotedblright\
of places where the gauge condition (\ref{gc}) is not fulfilled. Actually
such deviations from the vacuum inspired \textquotedblright
gauge\textquotedblright\ would mean that the current $j_{A}^{\nu }$ were not
zero but flowed in the direction of the $A^{\nu }$ field. In the coordinate
system in which the vacuum $A^{\nu }$ had only the time component $A^{0}$
this would mean \textquotedblleft fossil charges\textquotedblright\ at rest
forever. Having such charges at one's disposal the physical Lorentz
violation would actually be accessible. In fact the very finding of some
\textquotedblleft resting\textquotedblright\ charge of this type, which is
only able to follow the direction of the essentially vacuum four vector
field $A^{\nu }$, would by itself constitute a strong breaking of Lorentz
invariance: these charges do not move at all in the special frame. They
could be observed by means of the Coulomb fields, say, by which they would
be surrounded. But it should be stressed that we proved above that these
fossil resting charges cannot be produced, if the condition (\ref{gc}) is
fulfilled all over the Universe and there are no charges of this type to
begin with. They can only come from the start, but cannot be produced.

For simplicity we treated above only the case that the vacuum expectation
value of the $A^{\mu }$ field was pointed out in a timelike direction.
Actually, we mainly used that case to ensure that the curves having the $%
A^{\mu }$ as tangents in space-time would come from the past and run into
future, so that an assumption about the partial derivative $\frac{\partial
{\cal L}_{br}}{\partial A^{2}}$ being zero in the early Universe were
sufficient to ensure it once we had proved that it did not vary along these
curves. Note however that the frames in which the assumed initial conditions
$\frac{\partial {\cal L}_{br}}{\partial A^{2}}=0$ are satisfied actually
form a special non-Lorentz invariant set for the case of a space-like SLIV ($%
<A^{\mu }>$ $=n^{\mu }V$ with $n^{2}=-1$). In such a frame, a curve having a
corresponding space-like vacuum expectation value $<A^{\mu }>$ as a tangent
will generically have a component along the time direction associated with a
non-zero $<A^{0}>$. Thus the proof of our theorem follows in the same way
for the space-like as in the time-like case.

\section{Counter examples by initial conditions}

The obvious question to ask is whether we can make the SLIV become
observable by not having the initial conditions with $\frac{\partial {\cal L}%
_{br}}{\partial A^{2}}=0$ which was needed in the theorem. So let us think
about rather small deviations from this condition having a perturbatively
small $\frac{\partial {\cal L}_{br}}{\partial A^{2}}$. Even in this case the
conservation of the current $j_{A}^{\mu }$ following from the basic equation
(\ref{gme}) would enforce the partial derivative $\frac{\partial {\cal L}%
_{br}}{\partial A^{2}}$ to remain approximatively constant along the curves
tangential to the $A_{\mu }$ field. So we would indeed find that such
perturbatively small deviations from zero would remain constant along these
curves. In the approximate vacuum situation with a background vacuum with $%
A_{\mu }\approx V_{\mu }$ being constant in space-time ($V_{\mu }=n_{\mu }V$
where $V$ is a proposed SLIV scale) we could, for instance, in the $V_{\mu }$
timelike case talk about the frame in which the spatial components of $%
V_{\mu }$ are zero as the frame specified by $V_{\mu }$ and we would simply
have the curves tangential to $A^{\mu }$ or approximately to $V^{\mu }$ are
the timelike curves meaning the time tracks of resting particles. In this
situation the partial derivative $\frac{\partial {\cal L}_{br}}{\partial
A^{2}}$ will stay approximately constant as function of time, but could
(while we still consider it small) vary as a function of space. In the first
approximation we see from the expression (\ref{ja}) that $j_{A}^{\mu }$ now
represents a charge distribution $V^{0}\frac{\partial {\cal L}_{br}}{%
\partial A^{2}}$ which does not change with time. This charge distribution
is via the Maxwell equations - our equations of motion derived by the
variation of $A_{\mu }$ - observable, as are usual charges, e.g. via the
Coulomb field they cause. But now since $A_{\mu }$ or $V_{\mu }$ is neither
gauge nor Lorentz invariant in the SLIV case such observation of these
fields via the Maxwell equations means that these symmetries are effectively
broken and, therefore, SLIV becomes physically observable. Thus our theorem
does not work if the initial condition that the partial derivative $\frac{%
\partial {\cal L}_{br}}{\partial A^{2}}$ be zero is not fulfilled. So, this
condition mentioned in the theorem is quite needed.

We should really think of the deviations from zero of this $\frac{\partial
{\cal L}_{br}}{\partial A^{2}}$ as representing charge density that must be
fossil in the sense that we could not obtain it unless it were there to
begin with. One would therefore be reasonable to call the charge density $%
\rho _{fossil}=V^{0}\frac{\partial {\cal L}_{br}}{\partial A^{2}}$ a fossil
charge.We also remark that it does not change as time goes on and thus just
rests where it is to begin with. So we could call it {\em the fossil resting
charge density}. It is just via this resting fossil charge density that the
Lorentz non-invariance can come into the theories which in other respects
obey the conditions of our theorem but only lacks to obey the initial
condition requirements.

One should of course seek experimentally to look for such resting fossil
charges. If one had single quanta of them, one might observe them as
particles moving with a remarkably fixed velocity. If this velocity were
larger than the velocity of an atomic electron, such a fossil charge would
produce ionization tracks over long distances in an emulsion without
stopping. This would be a characteristic signal and its non-observation puts
a severe upper bound on the density of resting fossil charges with such a
velocity. However one might guess that their velocity would follow the
cosmic microwave background. In which case the velocity observed in practice
would be that of the earth relative to this microwave background, which is
of the order of 400 km/s \cite{pdg}. Unfortunately this velocity is too
small to produce ionization and the resting fossil charges would not be so
easily observed.

In the case of a space-like $<A^{\mu }>$, the fossil charges would look like
tachyons passing by. They would then give rise to Cerenkov radiation and
would also make tracks in emulsions. The non-observation of such effects
again places a severe upper bound on any fossil charge density.

\section{Counter example by allowing extra terms}

The assumptions made in our theorem above might seem a bit arbitrary and not
so simple, and indeed are not exceedingly beautiful. Apart from the
necessary requirement to have the gauge symmetric $F_{\mu \nu }^{2}$ form
for the vector field kinetic terms that we discuss in the next section, we
can wonder about why we could not let in the Lagrangian ${\cal L}_{br}$ the
\textquotedblleft gauge violating \textquotedblright\ high-dimension
operator terms which, for instance, could depend also on a extra
construction $A_{\mu }\overline{\psi }\gamma ^{\mu }\psi$ etc. in addition
to its dependence on $A^{2}$, $\overline{\psi }\psi$ and $|\phi |^{2}$. The
point, however, is that with such a type of term, say $|\phi |^{2}A_{\mu }%
\overline{\psi }\gamma ^{\mu }\psi$, would indeed lead to a model in which
the SLIV were observable - of course only through the non-renormalizable
terms that now had to be used. Nevertheless, it would in this case be a
genuine Lorentz non-invariant theory to live in, even if one starts with the
above SLIV vacuum. We can indeed see that if we allow the ${\cal L}_{br}$ to
depend also on $A^{\mu }\overline{\psi }\gamma _{\mu }\psi $ so as to become
a function of the type ${\cal L}_{br}(A^{2},A^{\mu }\overline{\psi }\gamma
_{\mu }\psi ,\overline{\psi }\psi ,|\phi |^{2})$ we get the new current $%
j_{A}^{\mu }$ coming from the variation of $A^{\mu }$ in the gauge symmetry
breaking part ${\cal L}_{br}$

\begin{equation}
j_{A}^{\mu }=A^{\mu }\frac{\partial {\cal L}_{br}}{\partial A^{2}}+\overline{%
\psi }\gamma ^{\mu }\psi \frac{\partial {\cal L}_{br}}{\partial (A^{\nu }%
\overline{\psi }\gamma _{\nu }\psi )}  \label{cur}
\end{equation}%
and thus the requirement of it being divergenceless $\partial _{\mu
}j_{A}^{\mu }=0$ would now no longer immediately lead to that the partial
derivative $\frac{\partial {\cal L}_{br}}{\partial A^{2}}$ would stay zero
along curves tangentially following the $A^{\mu }$ field even at first being
zero at such a curve in the early times. Rather we now have an extra term%
\begin{equation}
\overline{\psi }\gamma ^{\mu }\psi \frac{\partial {\cal L}_{br}}{\partial
(A^{\nu }\overline{\psi }\gamma _{\nu }\psi )}
\end{equation}
in the current $j_{A}^{\mu }$ (\ref{cur}) and when we require its
divergenceless $\partial _{\mu }j_{A}^{\mu }=0$ we get the divergence of
this extra term into the equation then obtained
\begin{equation}
(\partial _{\mu }A^{\mu })\frac{\partial {\cal L}_{br}}{\partial A^{2}}%
+A^{\mu }\partial _{\mu }\frac{\partial {\cal L}_{br}}{\partial A^{2}}+%
\overline{\psi }\gamma ^{\mu }\psi \partial _{\mu }\frac{\partial {\cal L}%
_{br}}{\partial (\overline{\psi }\gamma _{\nu }\psi A^{\nu })}+(\partial
_{\mu }\overline{\psi }\gamma ^{\mu }\psi )\frac{\partial {\cal L}_{br}}{%
\partial (\overline{\psi }\gamma ^{\nu }\psi A_{\nu })}=0  \label{div1}
\end{equation}%
This means that even if we started with the $\frac{\partial {\cal L}_{br}}{%
\partial A^{2}}$ equal to zero it would not stay zero because of the third
term in the equation (\ref{div1})
\begin{equation}
\overline{\psi }\gamma ^{\mu }\psi \partial _{\mu }\frac{\partial {\cal L}%
_{br}}{\partial (\overline{\psi }\gamma _{\nu }\psi A^{\nu })}\text{ \ }
\label{div2}
\end{equation}%
while the last term in it vanishes because the fermion current $\overline{%
\psi }\gamma ^{\mu }\psi $ is conserved. However, provided we arrange for a
fermion current flow in the direction of a non-zero gradient multiplier
being in the term (\ref{div2}), then the $\frac{\partial {\cal L}_{br}}{%
\partial A^{2}}$ will vary even if it were initially zero. In fact, since
the term $(\partial _{\mu }A^{\mu })\frac{\partial {\cal L}_{br}}{\partial
A^{2}}$ in the equation (\ref{div1}) is initially zero, there is no way to
avoid the variation of the partial derivative $\frac{\partial {\cal L}_{br}}{%
\partial A^{2}}$ along the curves having the $A^{\mu }$ fields as tangents
to be non-zero (see in the above).

Remember that we think of deviations of the $\frac{\partial {\cal L}_{br}}{%
\partial A^{2}}$ from zero as representing \textquotedblleft
fossil\textquotedblright\ charge.  Thus, in the above case, fossil resting
charge can indeed be produced by a suitable field configuration even if
there were no fossil charges to begin with. So, with the extra dependence of
the ${\cal L}_{br}$ on the $\overline{\psi }\gamma ^{\mu }\psi A_{\mu }$,
the SLIV should in principle be observable even in a world that had started
out without any so-called fossil charges.

\section{Why partial gauge symmetry?}

A crucial assumption in our theorem, that was designated as the partial
gauge invariance, is that the kinetic term for the vector field $A_{\mu }$
should take the standard gauge symmetric $F_{\mu \nu }^{2}$ form being also
(at least approximately) stable against radiative corrections stemming from
the gauge breaking terms ${\cal L}_{br}$ (\ref{gf}). Although we are not
able to single out this $F_{\mu \nu }^{2}$ form on any symmetry ground there
are two serious arguments in favor of such a choice.

The first argument is related to the physical requirement that the kinetic
part of the Hamiltonian should be bounded from below. A priori, one should
consider, instead of the $F_{\mu \nu }^{2}$, kinetic terms of the general
form, $a(\partial _{\mu }A_{\nu })^{2}+b(\partial _{\mu }A^{\mu })^{2}$,
where $a$ and $b$ are arbitrary constants. Note then that the first term in
this form looks like as the kinetic terms of four scalar fields with an
exception that the $A_{0}$ component has the \ \textquotedblleft
wrong\textquotedblright\ metric, if we choose $a$ negative so that the
spatial components get the right sign for getting positive energy density.
So, the $A_{0}$ containing term always gives a negative contribution to the
Hamiltonian unless there is a way to get rid of the $\partial _{0}A_{0}$
term from the Lagrangian and the spatial derivative on $A_{0}$ terms $%
(\partial _{i}A_{0})^{2}$, which also comes with the \textquotedblleft
wrong\textquotedblright\ sign in our term with coefficient $a$. For the
momentum $\Pi ^{\mu }(x)$ conjugate to $A_{\mu }(x)$,
\begin{equation}
\Pi ^{\mu }=2(a\partial ^{0}A^{\mu }+bg^{0\mu }\partial _{\nu }A^{\nu })
\end{equation}%
the Hamiltonian density becomes in a usual way just

\begin{equation}
{\cal H}(x)=a[(\partial _{0}A_{\mu })^{2}+(\nabla A_{\mu })^{2}]+b[(\partial
_{0}A_{0})^{2}-(\nabla \cdot \vec{A})^{2}]  \nonumber
\end{equation}%
which, for $b\neq -a$, can be written in the form
\begin{equation}
{\cal H}(x)=\frac{(\Pi _{0}+2b\nabla \cdot \vec{A})^{2}}{a+b}-a\vec{\Pi}%
^{2}+a(\nabla \cdot A_{0})^{2}-a(\nabla \vec{A})^{2}-b(\nabla \cdot \vec{A}%
)^{2}
\end{equation}
So, the only Lorentz covariant way to get rid of the $(\partial
_{0}A_{0})^{2}$ is to have the equality $b=-a$ . Alternatively we can get
this kinetic $A_{0}$ term with the right sign by choosing $b$ bigger than
the positive $-a$ in the above form. But the gradient energy term from $A_{0}
$, i.e. $(\nabla A_{0})^{2}$ still would come with the negative coefficient $%
a$  and thus formally there would be no bottom to the Hamiltonian anyway.
Now the equation of motion for $A_{0}$ may be written
\begin{equation}
(a+b)\partial _{0}\partial ^{0}A_{0}-a\bigtriangleup A_{0}-b\partial
_{0}\nabla \cdot \vec{A}=0
\end{equation}%
In the limit $a=-b$ this equation reduces to a constraint

\begin{equation}
-a\bigtriangleup A_{0}-b\partial _{0}\nabla \cdot \vec{A}=0
\end{equation}%
which is in fact the usual Gauss constraint $\nabla \cdot \overrightarrow{E}%
=0$. After partial integration in the action and throwing away boundary
terms, this case $a=-b$ leads to the standard kinetic term. Now it is
well-known from conventional QED that, if this Gauss constraint is inserted
into the Hamiltonian, positivity of the kinetic energy can be guaranteed.
However, there is no such constraint available for the general form of the
kinetic term. Thus boundedness from below of the Hamiltonian requires a
kinetic term of the standard gauge invariant $F_{\mu \nu }^{2}$ form.

There is another, more phenomenological argument for the absence of a gauge
non-invariant addition to the vector field kinetic term. This is that the
presence of the gauge non-invariant addition in the vector field kinetic term

\begin{equation}
{\cal L}_{kin}=-\frac{1}{4}F_{\mu \nu }^{2}-\frac{\beta }{2}(\partial _{\mu
}A^{\mu })^{2}  \label{kin}
\end{equation}%
(where $\beta $ is some positive constant) immediately leads to the domain
wall solution for the SLIV, caused by a standard quartic polynomial in $%
A^{2} $ contained in ${\cal L}_{br}$ (\ref{gf})
\begin{equation}
P(A^{2})=n^{2}\frac{M^{2}}{2}A_{\mu }^{2}-\frac{\lambda }{4}(A_{\mu
}^{2})^{2},\text{ }M^{2}>0,\text{ \ }n^{2}=n_{\mu }n^{\mu }=\pm 1\text{ \ }
\end{equation}%
where the vector field mass squared $n^{2}M^{2}$ may be positive or negative
for time-like or space-like Lorentz violation, respectively. Actually, one
can find that the vector field $A_{\mu }$ develops a domain wall solutions $%
V(n\cdot x)$

\begin{equation}
A_{\mu }=a_{\mu }+n_{\mu }V(n\cdot x),\text{ \ }n\cdot x=n_{\mu }x^{\mu }
\label{vev}
\end{equation}%
\[
V(n\cdot x)=\sqrt{\frac{M^{2}}{\lambda }}\tanh \left[ \frac{(n\cdot x)M}{%
\sqrt{2\beta }}\right]
\]%
with walls being oriented orthogonal to the SLIV direction $n_{\mu }$. \ \
Indeed, it is clear that the standard kinetic $F_{\mu \nu }^{2}$ term in the
${\cal L}_{kin}$ (\ref{kin}) is invariant under substitution (\ref{vev}),
while the second term in it leads to the differential equation for $V(n\cdot
x)$
\begin{equation}
\beta (\partial \cdot n)^{2}V+M^{2}V-\lambda V^{3}=0
\end{equation}%
whose solution is just given by Eq.(\ref{vev}). Note that these domain walls
are topologically stable since for both time-like SLIV (kink in a time) and
space-like SLIV (kink in a preferred space direction) cases the system
possesses the disconnected vacua with expectation values $\pm \sqrt{\frac{%
M^{2}}{\lambda }}$, respectively. Actually, any rotation from one vacuum to
another would necessarilly intersect the wall.

These walls could unavoidably lead to a wall-dominated Universe in the early
times and, therefore, to its immediate collapse. Thus we have to exclude the
presence of the corresponding gauge non-invariant addition to the $F_{\mu
\nu }^{2}$ kinetic term (taking constant $\beta =0$ in the ${\cal L}_{kin}$ (%
\ref{kin})) in this case on phenomenological grounds.

\section{Conclusion}

We have discusssed an influence of the spontaneous Lorentz violation on the
origin of the gauge internal symmetry in the general Abelian vector field
theory. We formulated the theorem which seems to govern this mechanism and
found the conditions under which the SLIV becomes physically observable in
terms of some primordially existing or created \textquotedblleft
fossil\textquotedblright\ charges with the property moving through the
Universe with a fixed velocity. The partial gauge symmetry, namely gauge
symmetry of the derivative containing terms in the Lagrangian was found to
be crucial for our consideration, first of all the gauge symmetry for vector
field kinetic term to avoid the botomlessness of the Hamiltonian and exclude
the domain wall solution in the SLIV theory that would otherwise lead to the
wall-dominated Universe in the early times and, therefore, to its immediate
collapse. As a result with gauge invariant kinetic term, one comes in the
minimal theory to the exactly QED case with the non-covariant gauge choice
as the only SLIV effect.

\section*{Acknowledgments}

We would like to thank Rabi Mohapatra for useful discussions. One of us
(C.D.F.) wants to thank the Niels Bohr Fund for support to visit Niels Bohr
Institute.

\end{document}